# Uniaxial zero thermal expansion in low-cost $Mn_2OBO_3$ from 3.5 to 1250 K

Chi-Hung Lee,* Cheng-Yen Lin, and Guan-Yu Chen

Department of Applied Physics, Tunghai University, Taichung 407224, Taiwan

*E-mail: leech@thu.edu.tw

**Abstract**

Unique zero thermal expansion (ZTE) materials are valuable for use in precision instruments, including electronics, aerospace parts, and engines. However, most ZTE materials have a temperature range less than 1000 K under which they do not expand. In this study, we present a uniaxial ZTE in the low-cost $Mn_2OBO_3$ with a thermal expansion coefficient of -4(1)×$10^{-7}$ $K^{-1}$ along the $X_1$ principal axis from 3.5 to 1250 K. The monoclinic structure of $Mn_2OBO_3$ remains stable over the entire temperature range in ambient conditions. Considerable thermal contraction on the $BO_3$ trigonal planar and thermal expansion on the $MnO_6$ octahedra combine to produce uniaxial ZTE. Temperature-dependent Raman scattering reveals anharmonic low-frequency modes associated with $MnO_6$ Rigid Unit Modes (RUMs), which likely play a critical role in driving thermal contraction in the $BO_3$ trigonal planar. No charge order-disorder transition, which could cause thermal contraction, was observed up to 1250 K.

**Keywords:** zero thermal expansion, lattice anharmonicity, synchrotron X-ray diffraction, neutron diffraction, Raman scattering

## Introduction

Positive thermal expansion (PTE), the process in which a compound's lattices expand when heated, occurs in a majority of materials. Materials that undergo thermal contraction or negative thermal expansion (NTE) along a single[1-9] or several[10-16] directions have been extensively studied due to their unusual physical properties. However, certain materials do not expand or contract as the temperatures increase;[17-29] these rare materials instead yield a zero thermal expansion (ZTE) $|\alpha| < 10^{-6}$ $K^{-1}$.

ZTE materials do not undergo thermally induced stress, making them valuable for use in aerospace equipment,[30] engines.[31] For thin-film technologies, ZTE materials which maintain a constant film thickness are beneficial for electronic components,[32] and optical parts.[33] However, reducing thermal mismatch strain caused by the difference between the thermal expansion coefficients of the film and the substrate remains a challenge.[34] Uniaxial ZTEs[25] are particularly useful in thin-film technologies, in contrast to volumetric ZTEs. In a uniaxial ZTE film, in which the film thickness remains constant in relation to temperature, the PTE along the in-plane directions expand in conjunction with the substrate, eliminating mismatch stress and stabilizing the film at high temperatures, in contrast to volumetric ZTE films.

The temperature range under which a ZTE material can be used determines its suitability for use. Only a few compounds undergo non-PTE under temperature range over 1000 K. Of those compounds that do not undergo PTE over such wide temperature range, a majority, including $ZrW_2O_8$,[10] $(Sc,Y)_2W_3O_{12}$,[35] $ScF_3$,[13] $Ca(Zr,Hf)F_6$,[36] and $FeZr_2$,[3] undergo NTE. Notable NTE and ZTE materials are summarized in Table 1. Most ZTE materials are derived from modifications or adaptations of NTE parent compounds and do not exhibit ZTE over a wide temperature range. The only ZTE material with a temperature range exceeding 1000 K is the Al-doped $Sc_{2-x}Al_xW_3O_{12}$ at

x=0.5, and no parent compound of any material was found to exhibit ZTE throughout that temperature range. This ZTE results from NTE along the *a* and *c* directions and PTE along the *b* direction;[24] no ZTE occurred along a specific crystallographic direction. Moreover, most compounds contain $(CN)^-$, F, Sc, Y, Hf, or Zr, which are expensive and more environmentally concerning compared to Mn and B.

Homometallic warwickites such as $Fe_2OBO_3$ possess interesting electronic and magnetic properties and high levels of geometric frustration and has attracted considerable research attention.[37-39] One homometallic warwickite that has been studied less frequently is $Mn_2OBO_3$.[38-43] In this article, we report uniaxial ZTE in $Mn_2OBO_3$ with a thermal expansion coefficient of -4(1)×$10^{-7}$ $K^{-1}$ along the $X_1$ principal axis from 3.5 to 1250 K. Thermal expansion was examined using temperature-dependent high-resolution X-ray diffraction and neutron diffraction. The combination of the large NTE on the B–O coordination bond lengths and moderate PTE on the Mn–O bond results in ZTE along the [h00] direction. The contraction of the $BO_3$ trigonal planar structure arises from anharmonic vibrations, driven by three-phonon Umklapp scattering, within the $MnO_6$ Rigid Unit Modes (RUMs).

## Materials and Methods

Powdered $Mn_2OBO_3$ was obtained from a mixture of MnO (Alfa Aesar, ≥99.99%), $Mn_2O_3$ (Sigma Aldrich, ≥99%), and $H_3BO_3$ (Honeywell Fluka, ≥99.8%) with the molar ratio of 2:1:2.5. The powder was pressed into a pellet and heated in an alumina crucible at 900 °C for 24 h in air. The sample was reground and heated at the same temperature and period 3 times. The product sample phase purity was verified at room temperature using the Bruker D2 powder diffractometer with the CuKα x-ray. High-resolution temperature-dependent synchrotron X-ray diffraction (XRD) patterns were collected at the TPS19A (temperature range 100 to 440 K) and TLS01C2 (300 to 1250 K) beamlines at the National Synchrotron Radiation Research Center in Taiwan. The sample was put in a quartz capillary tube with a diameter of 0.2 mm. The incident wavelength was fixed at 0.61992 Å, giving an instrumental resolution of $\Delta 2\theta = 0.02^o$ and $0.06^o$ for the TPS19A and the TLS01C2, respectively. Temperatures were controlled using a liquid nitrogen cryostream and a hot air gas blower. Diffraction patterns below the $LN_2$ temperature were collected using high-resolution neutron powder diffractometer (NPD) at Echidna in ANSTO. The XRD and NPD patterns were analyzed by the Rietveld method,[44] employing the General Structure Analysis System (GSAS),[45] to identify structural parameters. Low-temperature Raman scattering experiments were performed using a 785 nm laser. The temperature was controlled within a range of 3 to 300 K using a closed-cycle cryostat.

## Results and discussion

The XRD patterns of $Mn_2OBO_3$ collected at room temperature are shown in Fig. 1(a). All diffraction peaks corresponded to a monoclinic $P2_1/n$ (No.14) structure with lattice constants of $a$ = 9.29020(5) Å, $b$ = 9.53764(6) Å, $c$ = 3.24408(3) Å, and $\beta$ = 90.740(1)$^o$, a finding consistent with similar structural parameters in the literature.[38] No impurity phase was detected. In the refined crystalline structure plotted in Fig. 1(b), two inequivalent sites for Mn ions, Mn(1) and Mn(2), can be observed. From the Bond-Valence Sum calculation (BVS),[46] Mn(1) and Mn(2) correspond to $Mn^{3+}$ and $Mn^{2+}$, respectively. The structure is composed of distorted $Mn^{3+}O_6$ and $Mn^{2+}O_6$ octahedra, indicating that charge ordering occurs at room temperature. A distorted $BO_3$ trigonal planar, composed of O(2), O(3), O(4), is located in the largest void between the $MnO_6$ octahedra.

Temperature dependent synchrotron X-ray and the neutron diffraction patterns are shown in Fig. 2. Surprisingly, the position of the {200} reflection was unchanged, indicating that thermal expansion is not significant. The other reflection at {020} shifted to a lower angle at high temperatures, indicating the occurrence of PTE along this direction. The thermal expansions were examined using neutron diffraction below 60 K. Unfortunately, the structure factor of {h00} reflections by neutron scattering are too small to be detected. However, the positions of reflections with high h components, such as {520}, appear unchanged, while the {012} reflections shift slightly to lower angles as the temperature increases, Fig. 2(c). This suggests that thermal expansion is less pronounced along the h direction compared to the k and l directions.

Lattice parameters obtained from the fits to the XRD and NPD patterns obtained using Rietveld refinement are plotted in Fig. 3. The overall thermal expansion along the *a* direction was much less significant than expansion along the *b* and *c* directions. However, due to the slight deviation of the beta angle from 90°, the lattice constant *a* alone is insufficient to accurately describe the thermal expansion. The thermal expansion coefficients along the principal axes ($X_1$, $X_2$, $X_3$) were determined using Principal Axis Strain Calculator (PASCal) software, [47, 48] by analyzing the eigenvalues and eigenvectors of the thermal expansion tensor. The thermal expansion coefficients along the principal axes are summarized in Table 2. The ZTE was observed along $X_1$ direction with the α of $-4(1)\times10^{-7}$ $K^{-1}$. The $X_1$ is slightly offset from the *a*-axis by an angle of $1.04^o$. The $X_2$ direction had a low α of $8.0(4)\times10^{-6}$ $K^{-1}$, a coefficient slightly too large to address ZTE, whereas PTE was notable in the $X_3$ direction. The ZTE along the $X_1$ direction and the slight PTE along the $X_2$ direction resulted in a slight volumetric thermal expansion of only $2.3(9) \times 10^{-5}$ $K^{-1}$. Figure S1 in the Supplementary materials displays the observed and calculated diffraction patterns at 3.5 and 1250 K and indicates that no impurity phases or structural transitions developed.

Lattice ZTE from 300 to 1250 K was confirmed in a second sample, as shown in Fig. S2 in the Supplementary materials.

Significant shrinkage of the $BO_3$ trigonal planar in $Mn_2OBO_3$ was observed, potentially serving as the origin of ZTE. Figure 4(a) shows the thermal variations of the average B–O, $Mn^{3+}$–O, and $Mn^{2+}$–O bond lengths obtained from Rietveld refinement to the high-resolution synchrotron XRD patterns. It is important to note that the bond lengths calculated from atomic coordinates are referred to as coordination bond lengths. The distinction between coordination bond lengths and real bond lengths will be discussed later. The average B–O bond underwent a large thermal contraction with a α of $-1.8 \times 10^{-5}$ $K^{-1}$. The average $Mn^{3+}$–O bond exhibited a small PTE with a α of $5.5 \times 10^{-6}$ $K^{-1}$, whereas the average $Mn^{2+}$–O bonds exhibited PTE with a coefficient in the order of $10^{-5}$ $K^{-1}$. The uniaxial ZTE is attributed to the significant thermal contraction of the B-O bond combined with the minimal thermal expansion of the $Mn^{3+}$–O bond. Complicated thermal behaviors along the B–O(2), B–O(3), and B–O(4) bonds are shown in Fig. 4(b). In addition to the contraction on the $BO_3$ trigonal planar, a decrease in the O(2)–B–O(3) and O(3)–B–O(4) bond angles and an increase in the O(2)–B–O(4) bond angles while warming resulted in the less-distorted $BO_3$ trigonal planar, as shown in Fig. 4(c).

This contraction and expansion along the edges of the $MnO_6$ octahedra and the $BO_3$ trigonal planar are shown in Fig. 5(a). As temperature increases, the decrease in the O(2)–B–O(3) angle and contraction of the B–O(2) bond shorten the O(2)–O(3) distance, primarily along the *a* and $X_1$ directions. Along the $X_2$ direction, the increase in the O(2)–B–O(4) angle and decrease in the O(3)–B–O(4) angle expand the lattice, while the contraction of the B–O(4) bond counteracts this expansion, resulting in slight PTE along the $X_2$ direction. Combine with the thermal expansion of the $MnO_6$ octahedra, this results in ZTE along the $X_1$ direction and slight PTE along the $X_2$ direction.

Since the $BO_3$ trigonal planar lies near the $X_1$- $X_2$ plane (close to *a*-*b* plane), a normal PTE was observed in the $X_3$ direction. Significant anisotropic electronic charge redistribution around the B, O(2), and O(4) ions is observed. Electronic charge densities were obtained using the GSAS program[45] by refining the X-ray diffraction profiles. The difference in electronic charge densities between 440 K and 110 K is plotted in Fig. 5(b). Positive density indicates increased electron density at 440 K. The anisotropic shifts of electrons surrounding the B, O(2), and O(4) ions suggest ion movements, resulting in a widening of the O(2)–B–O(4) angle, a reduction of the O(2)–B–O(3) and O(3)–B–O(4) angles, and a shortening of the B–O(2) and B–O(4) bonds. The O(3) ions exhibit a typical thermal redistribution, where electrons becoming more delocalized at higher temperatures.

In an ideal harmonic lattice vibration, no thermal expansion or contraction occurs, and the Raman frequency remains constant with temperature. Lattice anharmonicity, where non-linear interactions between atoms lead to structural adjustments, can account for lattice expansion or contraction[49, 50]. Figures 6(a) and (b) show two representative peaks at 154 $cm^{-1}$ (denoted as P1) and 1150 $cm^{-1}$ (denoted as P2) of $Mn_2OBO_3$ at 3 and 300 K. The P1 mode, shifted to the Mn sites due to $Ca^{2+}$ doping (Fig. S3(f) in the Supplementary materials), indicates that this mode is associated with the $MnO_6$ octahedra. Low-frequency modes, typically below 300 $cm^{-1}$, are generally resulted from the vibrations of the Rigid Unit Modes (RUMs). The presence of the low-frequency P1 mode suggests a potential candidate for the vibrations associated with the RUMs of the $MnO_6$ octahedra, reflecting their flexibility and ability to tilt or rotate. As temperature increases, the softening of the P1 mode signals lattice anharmonicity (Fig. 6(c)), which results in asymmetric vibrations and indicates increased flexibility and greater freedom for tilting or rotation within the $MnO_6$ RUMs. This increase in flexibility induces strain or adjustments in the surrounding lattice, resulting in changes to bond lengths and angles within the $BO_3$ trigonal planar. On the other hand, the energy of the P2 mode shifts when the B is substituted by $^{11}B$, while the P1 mode remains unchanged, as shown in Figs. S3(d) and (e) in the Supplementary materials. This evidence the P2 mode is associated with the B–O bond vibration. Surprisingly, the P2 mode shifts to a higher frequency as the temperature increases, signaling a stiffening of the B–O bonds. The stiffening indicates bond contraction within the $BO_3$ units, where the increased thermal energy leads to subtle adjustments in the B–O bond lengths and angles (Fig. 4(b), (c)), resulting in a strengthening of the B–O bonds. Furthermore, the scattering rates ($1/\tau$), obtained from the width of the Raman peaks, for both P1 and P2 can described very well from 3.3 to 300 K by the three-phonon Umklapp scattering process model ($\omega^2 T e^{-C/T}$) proposed by Asen-Palmer et al[51] rather than the normal process scattering model ($\omega^2 T^3$ or $\omega^2 T^4$). Based on the selection rules of phonon scattering, long-wavelength longitudinal phonons are not allowed in the Umklapp process.[52] This suggests that the P1 and P2 modes are transverse phonons.

In the case of $ScF_3$, the shortening of Sc–Sc distances is resulted from the transverse vibrations of F ions due to the $ScF_6$ RUMs.[13] Similarly, in $Mn_2OBO_3$, the contraction could be driven by the transverse vibration of oxygen, which arises from interactions with $MnO_6$ RUMs. The thermal ellipsoids of O(2), O(3), and O(4) at 1250 K, derived from the anisotropic atomic displacement parameters $U_{ij}$, which represent the magnitudes and directions of thermal vibrations, are shown in Fig. 7. The $U_{ij}$, obtained from the Rietveld refinement, of each atom are summarized in Table S1 in the Supplementary materials. These ellipsoids indicate that oxygen ions exhibit larger vibrational amplitudes perpendicular to the B–O bond compared to parallel, indicating the presence of the transverse vibrations of the oxygens. In cases with significant transverse vibrations, the coordination bond length, representing the distance between the average atomic positions, deviates from the real bond length, as illustrated in Fig. S4 in Supplementary materials. At higher temperatures, the larger vibrational amplitude leads to a smaller coordination bond length. A similar phenomenon appears in $ScF_3$. As the temperature increases, the real Sc–F bond length expands while the Sc–Sc distance contracts.[13] Since the mean atomic position of F ion lies at the center of the Sc–Sc, this leads to an apparent shortening of the Sc–F coordination bond length.

ZTE and NTE, which conflict the normal physical behavior, may be driven by various mechanisms, such as electrons,[14-16, 53] and phonons.[3, 4, 10, 18-20] Phonon-driven ZTE and NTE can occur under a temperature range as high as several hundred K, whereas the electron-driven ZTE and NTE occur under a much narrower temperature range. In some materials, NTE was accompanied by a charge order-disorder transition. NTE occurred in $V_2OPO_4$ upon heating to the charge disorder state.[15] A charge order-disorder transition at 317 K also has been observed in the similar compound $Fe_2OBO_3$.[37] However, in $Mn_2OBO_3$, the BVS calculation[46] shows that Mn(1) and Mn(2) remain as $Mn^{3+}$ and $Mn^{2+}$, respectively, up to 1250 K (Fig. S5 in the Supplementary materials). No mix-valence such as $Mn^{2.5+}$ was present, indicating that the long-range charge order persists up to 1250 K. ZTE between 3.5 and 1250 K was found in the charge order state. Long-range charge order at a lower temperature of 773 K has been observed in another study,[43] which is consistent with our results. Phonon-driven NTE typically occurs in materials that have structures with high void volume.[4, 10, 18-20] As temperatures increase, the strengthening of transverse vibrations on the framework shortens the bond lengths, resulting in thermal contraction.[54] High void volume is thus necessary to produce a strong transverse vibration for significant NTE. The $Mn_2OBO_3$ have been used as cathodes and anodes in a lithium-ion battery,[55, 56] indicating that the void in $Mn_2OBO_3$ is large enough even for lithium-ions insertion. The $BO_3$ trigonal planar, lying near the $X_1$–$X_2$ plane, is located at the largest void between the $MnO_6$ octahedra and is allowed to vibrate transversely along the $X_3$ direction. The ZTE of the $Mn_2OBO_3$ is attributed to the contraction of the $BO_3$ trigonal planar due to the anharmonic vibrations of $MnO_6$ RUMs caused by three-phonon Umklapp scattering. Other NTE materials composed of $BO_3$ trigonal planar were also been studied. $Co_3BO_5$ is known as the electron-driven NTE, where a charge order-disorder transition is associated.[53] NTE in $(Ca,Sr)B_2O_4$ is proposed to be driven by low-frequency phonons, with thermal contraction occurring along the $BO_2$ chains instead of the $BO_3$ trigonal planar.[57] Both $Co_3BO_5$ and $(Ca,Sr)B_2O_4$ exhibit NTE within a temperature range no more than 300 K. In $Mn_2OBO_3$, the boron ions are separated by the $MnO_6$ octahedra, and no $BO_2$ chains are formed. The physical mechanisms of ZTE in $Mn_2OBO_3$ may be different from those in $Co_3BO_5$ and $(Ca,Sr)B_2O_4$.

## Conclusions

We reported a uniaxial zero thermal expansion (ZTE) material under temperatures over 1000 K. The low-cost and less environmentally concerning $Mn_2OBO_3$ exhibits a ZTE with an average thermal expansion coefficient of $-4(1)\times10^{-7}$ $K^{-1}$ along the $X_1$ principal axis from 3.5 to 1250 K. A significant thermal contraction on the B–O coordination bond length due to the anharmonic vibrations of $MnO_6$ Rigid Unit Modes (RUMs) caused by three-phonon Umklapp scattering was observed. ZTE in $Mn_2OBO_3$ is the result of the opposition of thermal contraction on the $BO_3$ trigonal planar and thermal expansion on the $MnO_6$ octahedra.

## Acknowledgements

The authors would like to thank the National Science and Technology Council of Taiwan for supporting this research through the grants NSTC-111-2112-M-029-010-MY3, 113-2112-M-008 -015, and 113-2112-M-029-006. The authors would like to thank Dr. Chin-Wei Wang, Dr. Po-Ya

Chang and Mr. Chung-Kai Chang of the National Synchrotron Radiation Research Center for x-ray and neutron beamtime assistance.

**Figures**

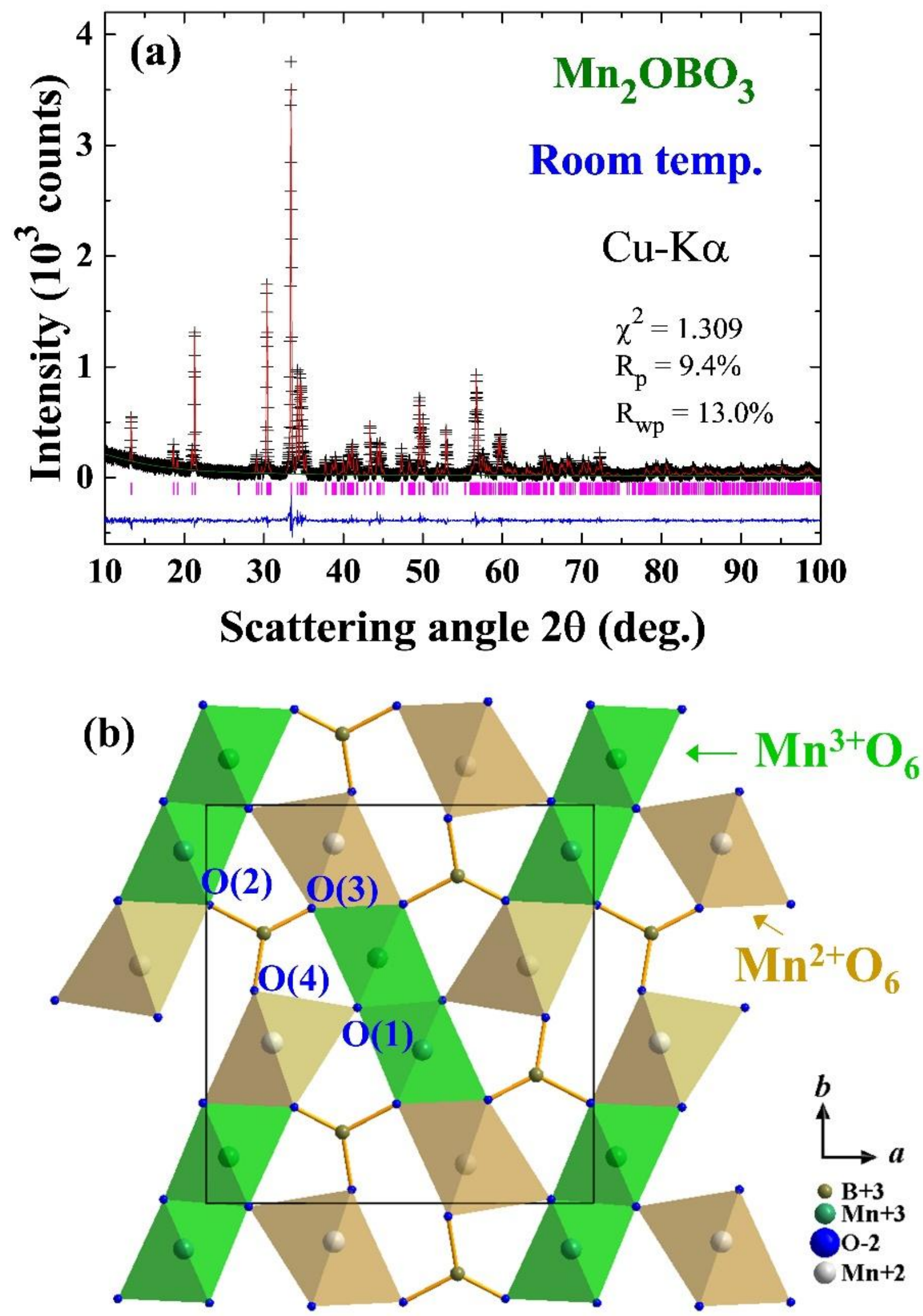

Figure 1. (a) Observed (crosses) and calculated (solid line) X-ray diffraction patterns at room temperature, assuming a monoclinic *P2₁/n* structure. The differences between the calculated and observed patterns are plotted at the bottom. (b) Schematic of the arrangement of the $Mn^{+3}O_6$ and $Mn^{2+}O_6$ octahedra and the $BO_3$ trigonal planar in $Mn_2BO_4$ projected into the *a-b* plane.

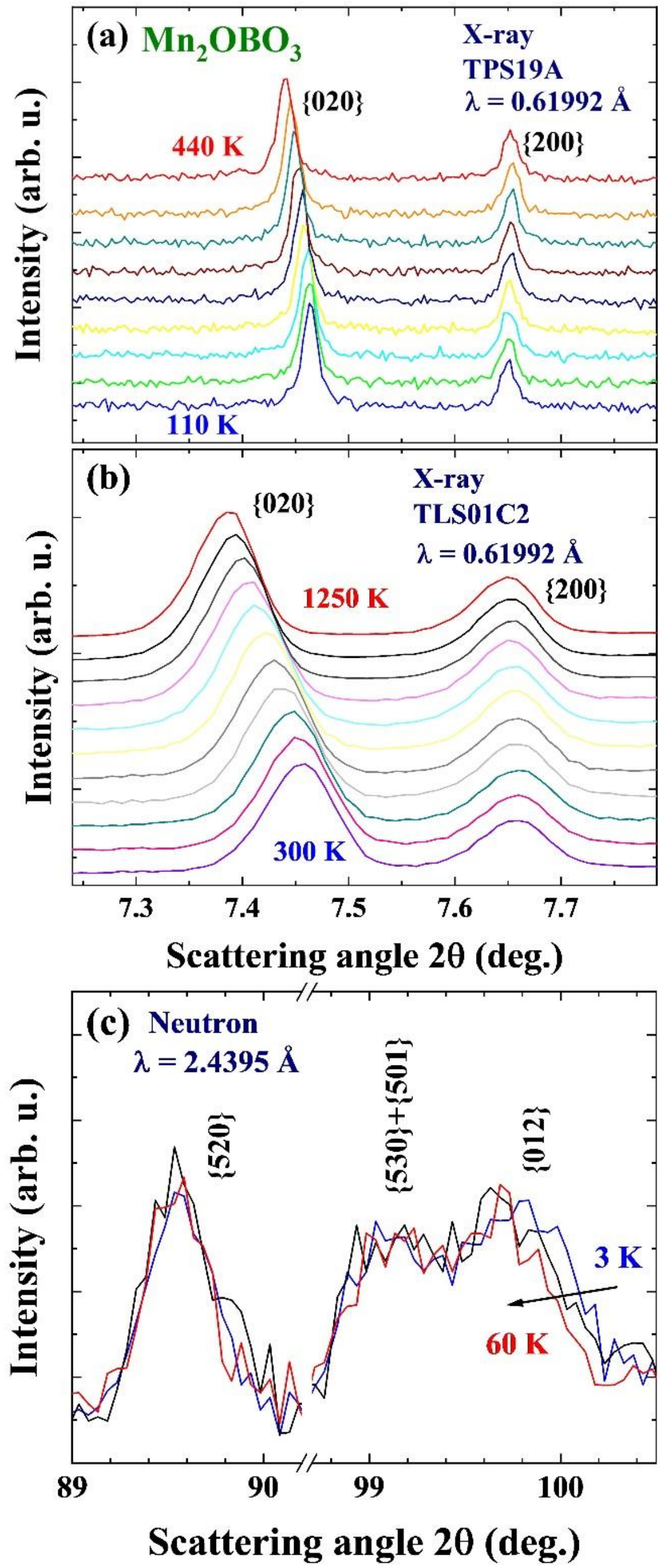

Figure 2. High-resolution synchrotron X-ray diffraction patterns collected at (a) TPS19A, (b) TLS01C2 in NSRRC and (c) neutron scattering at different temperatures. The position of the {200} and {520} reflections remain unchanged as the temperature increases.

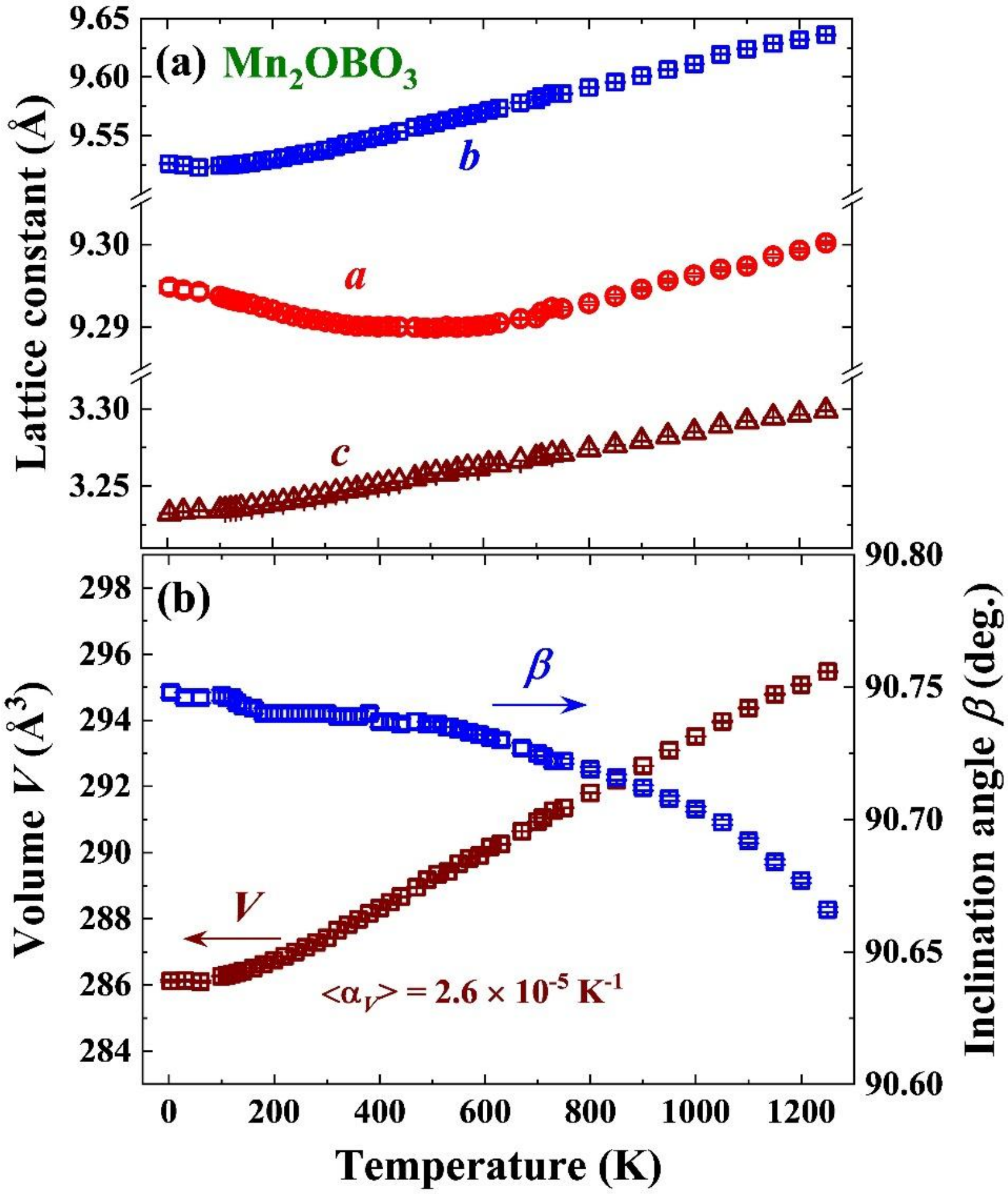

Figure 3. Temperature-dependent lattice parameters (a) *a*, *b*, and *c*, (b) $\beta$ and volume.

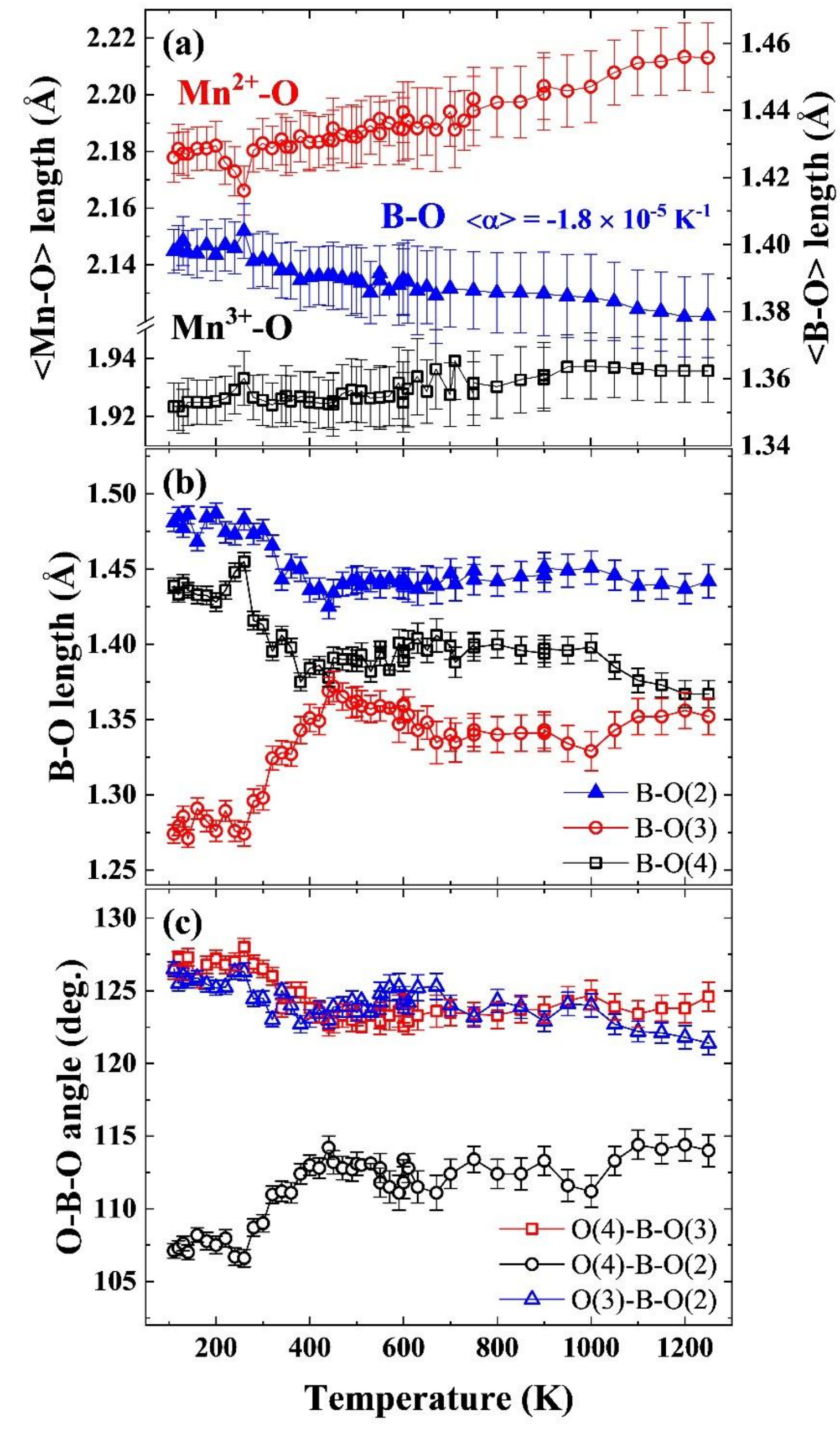

Figure 4. (a)Thermal variations in the average coordination bond lengths of $Mn^{3+}$–O, $Mn^{2+}$–O, B–O. Details of thermal variations in B–O (b) coordination bond lengths and (c) angles.

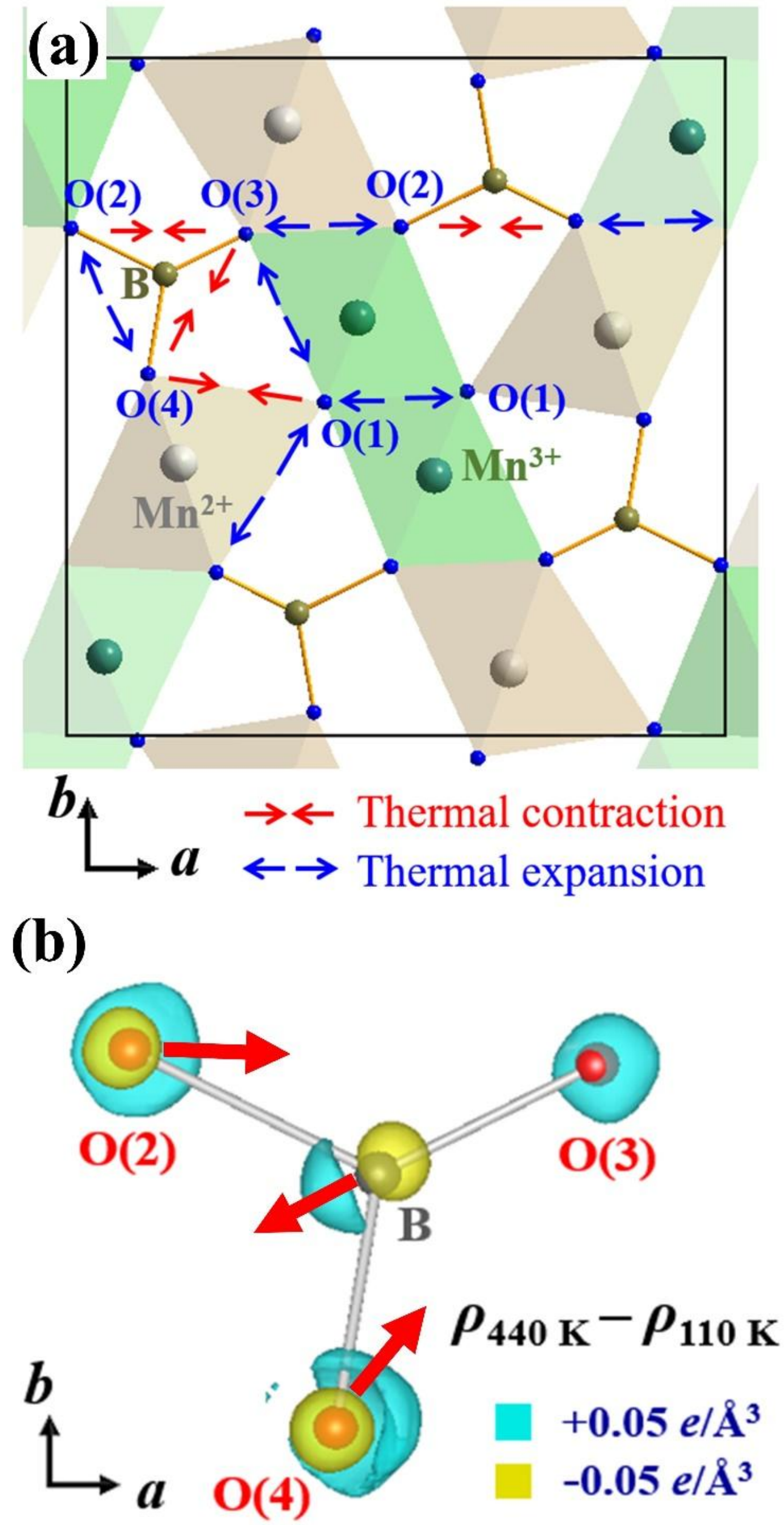

Figure 5. (a) Schematic of thermal contraction and expansion on the edge of the $MnO_6$ octahedron and the $BO_3$ trigonal planar. (b) Difference of electronic charge densities between 440 and 110 K.

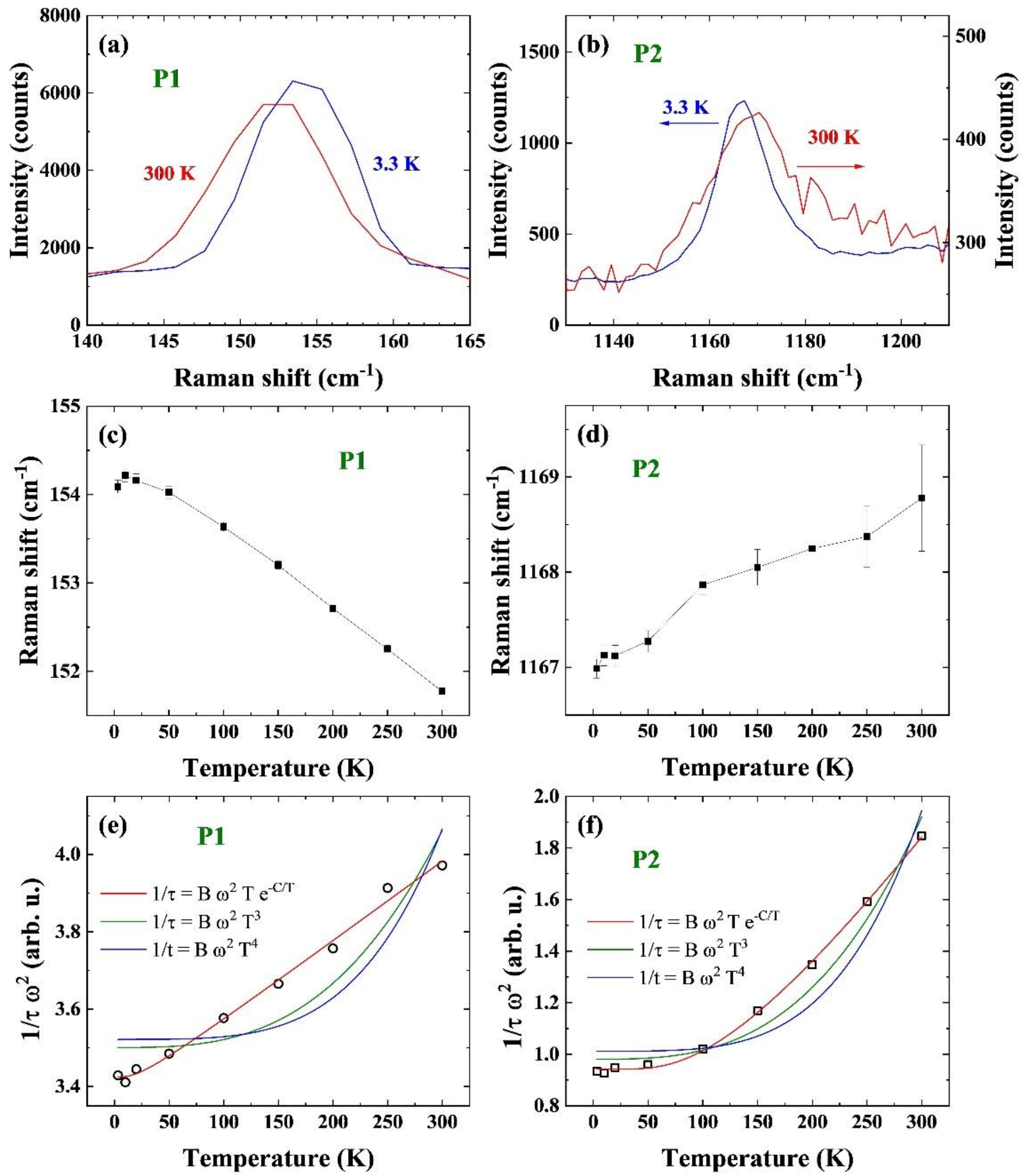

Figure 6. Raman scattering of (a) P1 and (b) P2 modes at 3.3 K and 300 K for direct comparison. Temperature dependence of the Raman peak positions for (c) P1 and (d) P2 modes. Temperature dependence of the phonon scattering rate ($1/\tau$) normalized by the square of the Raman frequency for (e) P1 and (f) P2 modes.

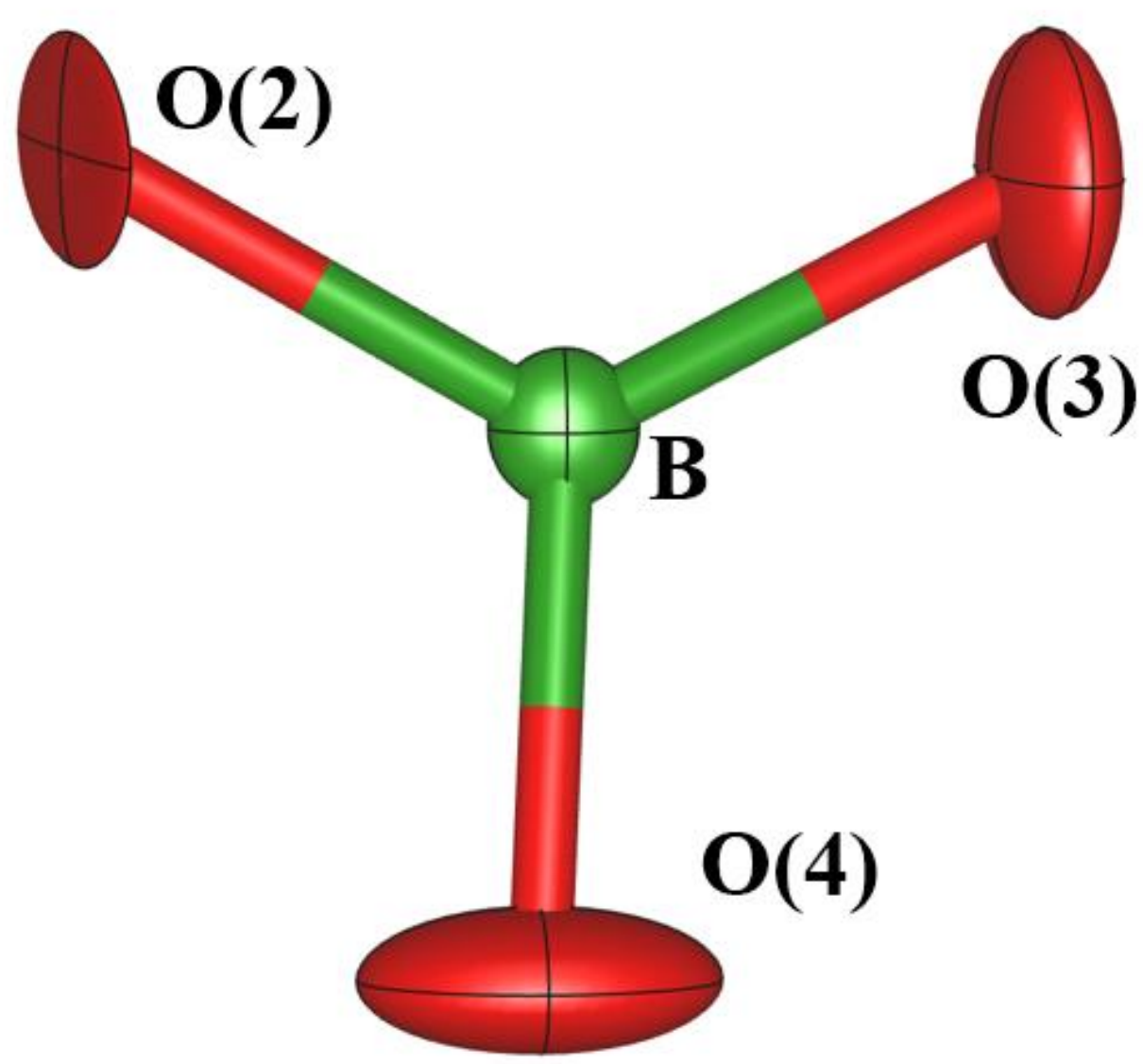

Figure 7. Thermal ellipsoid representation of the $BO_3$ trigonal planar at 1250 K. Significant thermal displacements perpendicular to the B–O bonds are observed.

**Table 1**. Summary of notable NTE and ZTE materials.

| NTE Material | α ($10^{-6}$ $K^{-1}$) | Temp. range (K) | Mechanism | Ref. |
|---|---|---|---|---|
| $ZrW_2O_8$ | $\alpha_a \approx -9$ | 0.3–1050 | RUMs | 10 |
| $Ca(Zr,Hf)F_6$ | $\alpha_l$ = -18 to -22 | 10–1022 | RUMs | 36 |
| $Y_2W_3O_{12}$ | $\alpha_l = -7$ | 15–1373 | RUMs | 35 |
| $Cu_2P_2O_7$ | $\alpha_v = -27$ | 5–375 | RUMs | 12 |
| $FeZr_2$ | $\alpha_c = -34$ | 93–1078 | RUMs | 3 |
| Polyarylamide (DBCOD) | $\alpha_c = -1200$ | 290–315 | Conformational subunit changes | 1 |
| $V_2OPO_4$ | $\alpha_v = -5.8$ | 620–750 | Charge ordering | 15 |
| $Ag_3[Co(CN)_6]$ | $\alpha_c$ = -120 to -130 | 16–500 | RUMs | 18 |
| $ScF_3$ | $\alpha_l = -14$ | 60–1100 | RUMs | 13 |
| ZTE Material | α ($10^{-6}$ $K^{-1}$) | Temp. range (K) | Mechanism | Ref. |
| $Li_{0.04}(Sc_{0.94}Fe_{0.06})F_3$ | $\alpha_l = -0.75$ | 150–425 | RUMs | 21 |
| $(Sc_{0.85}Ga_{0.05}Fe_{0.1})F_3$ | $\alpha_l = 0.23$ | 300–900 | RUMs | 29 |
| YbGaGe | Near-ZTE | 100–400 | Electronic transition | 17 |
| Prussian Blue Analogues | $\alpha_a = -1.47$ | 4.2–300 | RUMs | 19 |
| $PbTiO_3$ fibers | $\alpha_v = -1.41$ | 293–623 | Mesopores | 22 |
| $Tb(Co_{1.9}Fe_{0.1})$ | $\alpha_l = 0.48$ | 123–307 | Magneto-volume effect | 23 |
| $(Zr_{0.65}Nb_{0.35})_{1-y}Fe_yFe_2$ | $\alpha_a = 0.47$ | 4–425 | Magneto-volume effect | 26 |
| $Sc_{1.5}Al_{0.5}W_3O_{12}$ | $\alpha_v = -0.06$ | 4–1400 | Polyhedral distortion | 24 |
| $Ho_{0.04}Fe_{0.96}$ | $\alpha_a = 1.19$ | 100–335 | Dual-phase synergy | 25 |
| $Er_2Fe_{14}B$+α-Fe | $\alpha_a = -0.81$ | 100–550 | Dual-phase synergy | 27 |
| $(Ca,Sr)B_2O_4$ | $\alpha_l$ =-2.4 to +0.29 | 13–280 | RUMs | 57 |
| $Zn_4B_6O_{13}$ | $\alpha_a = 1$ | 13–273 | RUMs | 28 |

**Table 2.** Thermal expansion coefficients from 3.5 to 1250 K along three principal axes.

| Axes | $\alpha$ ($10^{-6}$ $K^{-1}$) | Directions | | |
|---|---|---|---|---|
| | | *a* | *b* | *c* |
| $X_1$ | -0.4(1) | 0.9998 | 0 | -0.0182 |
| $X_2$ | 8.0(4) | 0 | 1 | 0 |
| $X_3$ | 15.6(3) | 0.0067 | 0 | 1 |
| $V$ | 23(1) | | | |

## Supplementary materials

# Uniaxial zero thermal expansion in low-cost $Mn_2OBO_3$ from 3.5 to 1250 K

Chi-Hung Lee,* Cheng-Yen Lin, and Guan-Yu Chen

Department of Applied Physics, Tunghai University, Taichung 407224, Taiwan

*E-mail: leech@thu.edu.tw

### Thermal stability

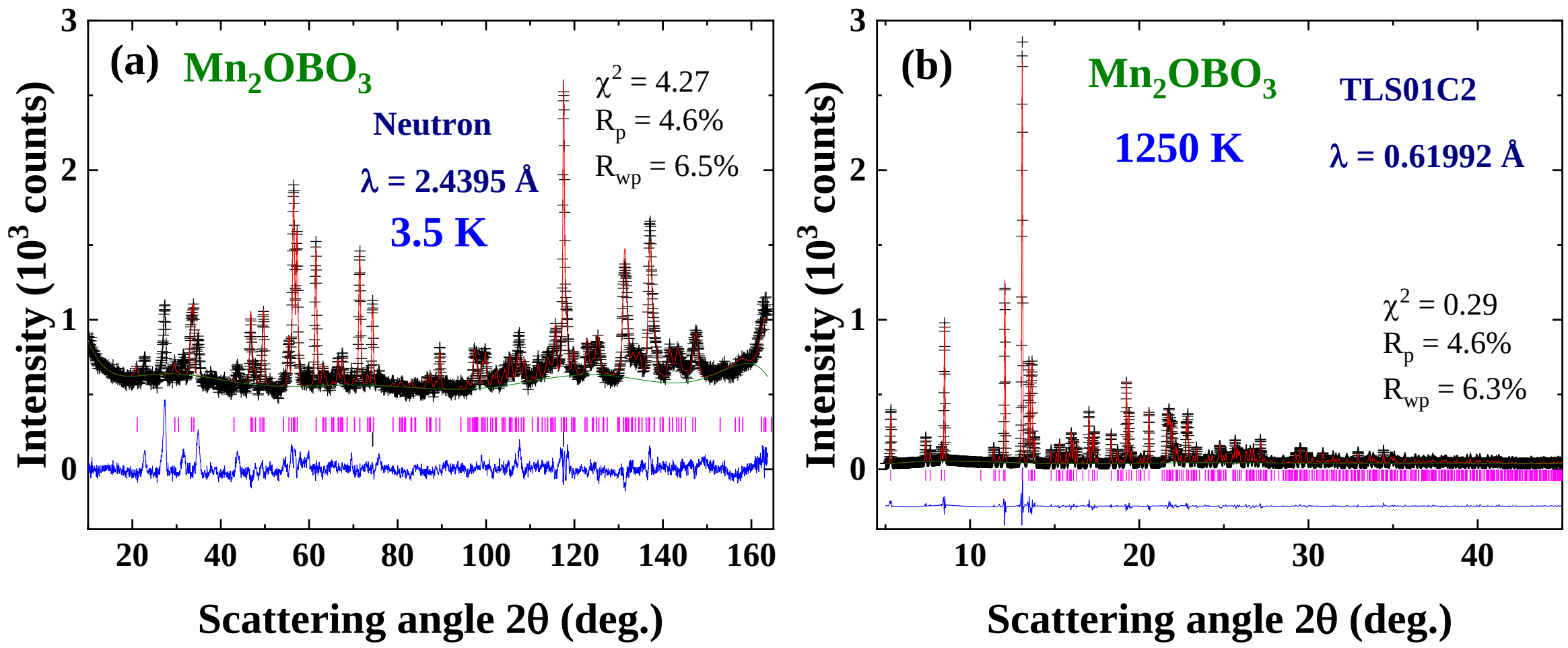

**Figure S1**. Observed (crosses) and calculated (solid line) (a) neutron diffraction pattern at 3.5K and (b)x-ray diffraction patterns at 1250 K, assuming a monoclinic P*2₁*/*n* structure, indicating the structure remains stable and no impurity phases or structural transition developed up to 1250 K. The undefined peaks at 3.5 K are magnetic diffraction peaks. The second phase at 3.5 K is the iron screw located in the sample holder.

## Confirm ZTE with a second sample (sample-B)

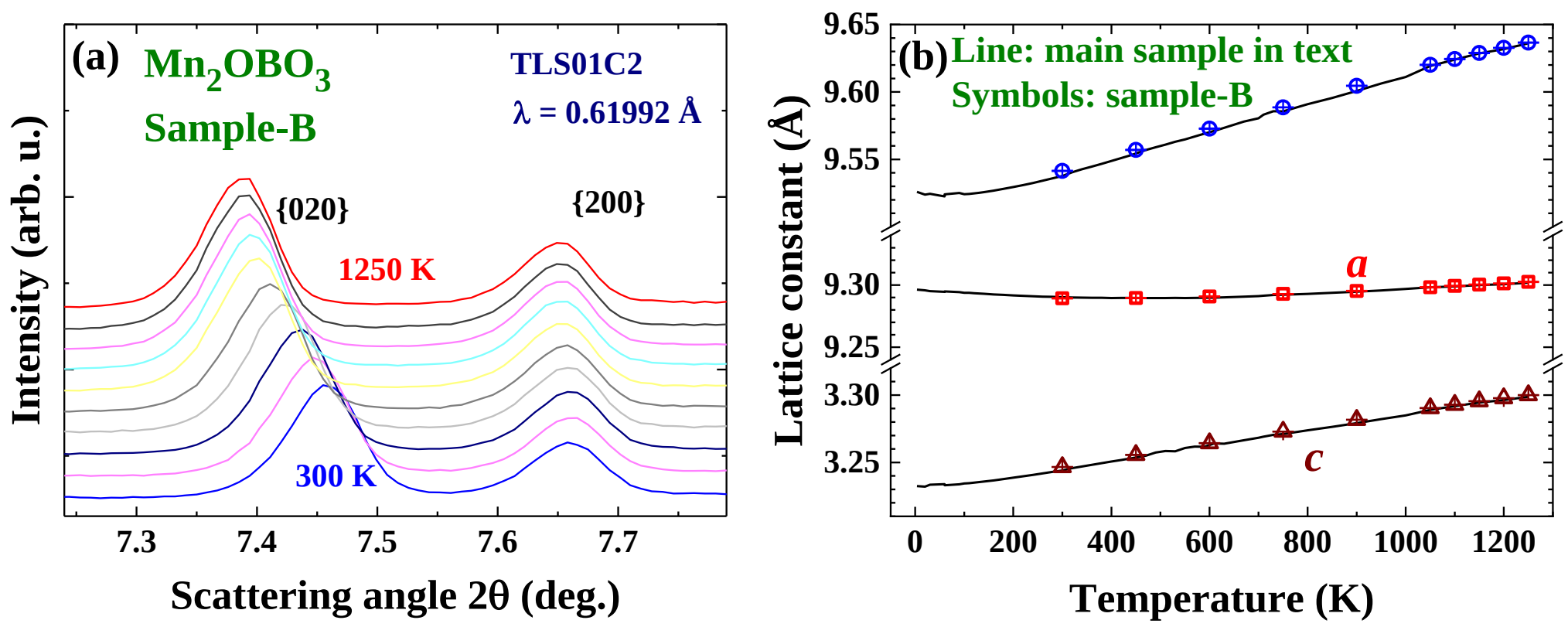

**Figure S2**. (a) High resolution synchrotron x-ray diffraction patterns of sample-B collected at TLS01C2 in NSRRC at different temperatures. The position of the {200}ffo reflection remains unchanged while warming. (b) Comparison on the lattice constants between sample-B (open symbols) and the main sample (solid lines).

## Raman scattering results

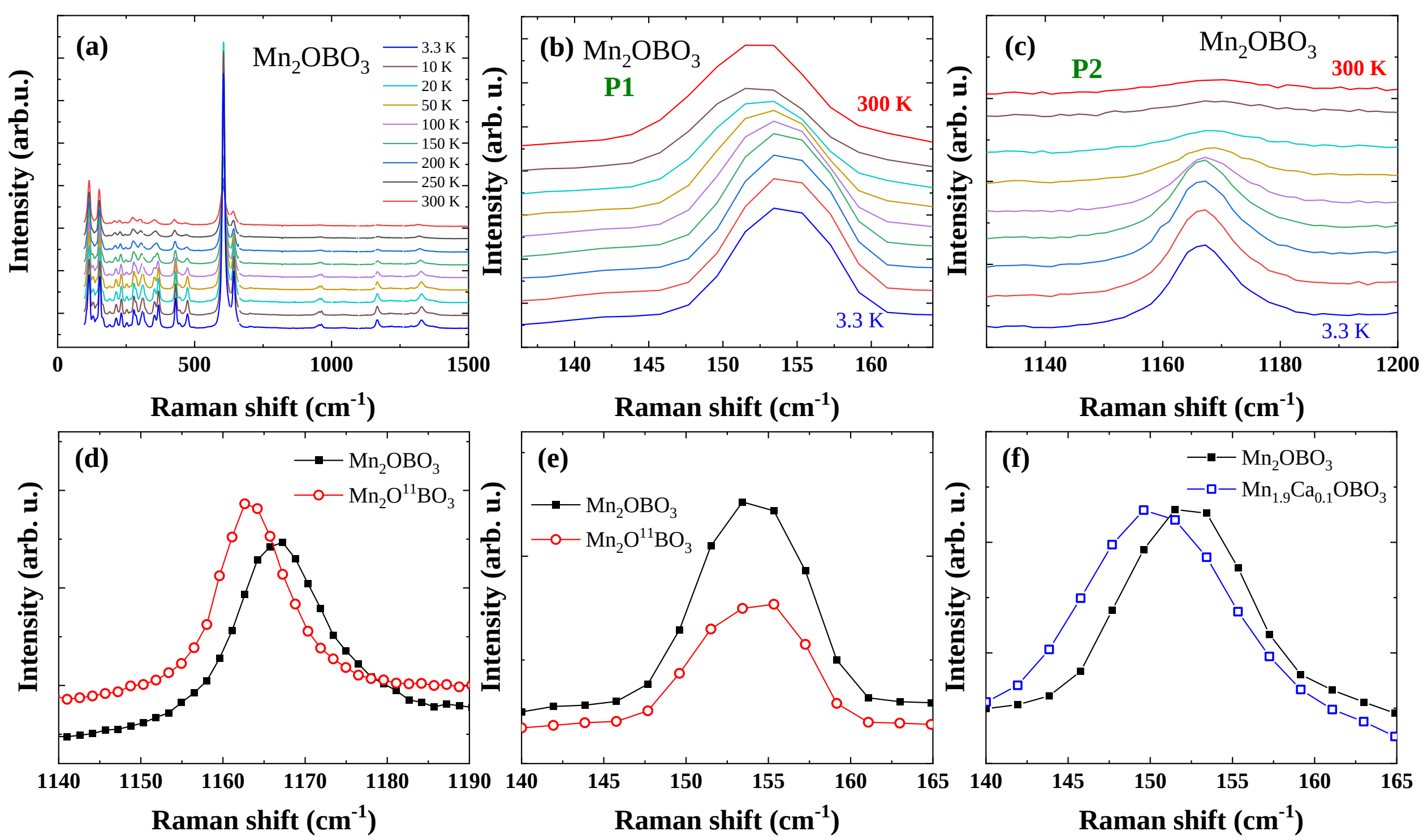

**Figure S3**. (a) Raman scattering spectra of $Mn_2OBO_3$ at various temperatures. (b) Temperature dependent of the P1 mode and (c) the P2 mode. (d) P2 and (e) P1 modes for $Mn_2OBO_3$ with $^{11}B$ isotope substitution. (f) P1 mode for $Ca^{2+}$-doped $Mn_2OBO_3$.

## Anisotropic atomic displacement parameters

**Table S1.** Anisotropic atomic displacement parameters $U_{ij}$ at 1250 K, with units in Å$^2$.

| Atom | 100*$U_{11}$ | 100*$U_{22}$ | 100*$U_{33}$ | 100*$U_{13}$ |
|---|---|---|---|---|
| B | 1.4(5) | 1.4(4) | 0.8 (5) | 0.5(4) |
| O(2) | 0.02(28) | 3.4(4) | 9.5(5) | 0.9(3) |
| O(3) | 0.9(3) | 6.1(4) | 5.8(4) | 0.6(3) |
| O(4) | 8.6(4) | 1.2(3) | 6.0(4) | 0.5(3) |
| O(1) | 1.7(3) | 3.7(3) | 5.1(4) | 0.3(3) |
| $Mn^{2+}$ | 4.09(9) | 4.49(7) | 3.47(9) | 0.43(6) |
| $Mn^{3+}$ | 1.92(7) | 2.08(7) | 2.16(7) | 0.44(5) |

$U_{12} = U_{23} = 0$, as constrained by the $P2_1/n$ symmetry.

## Coordination bond length

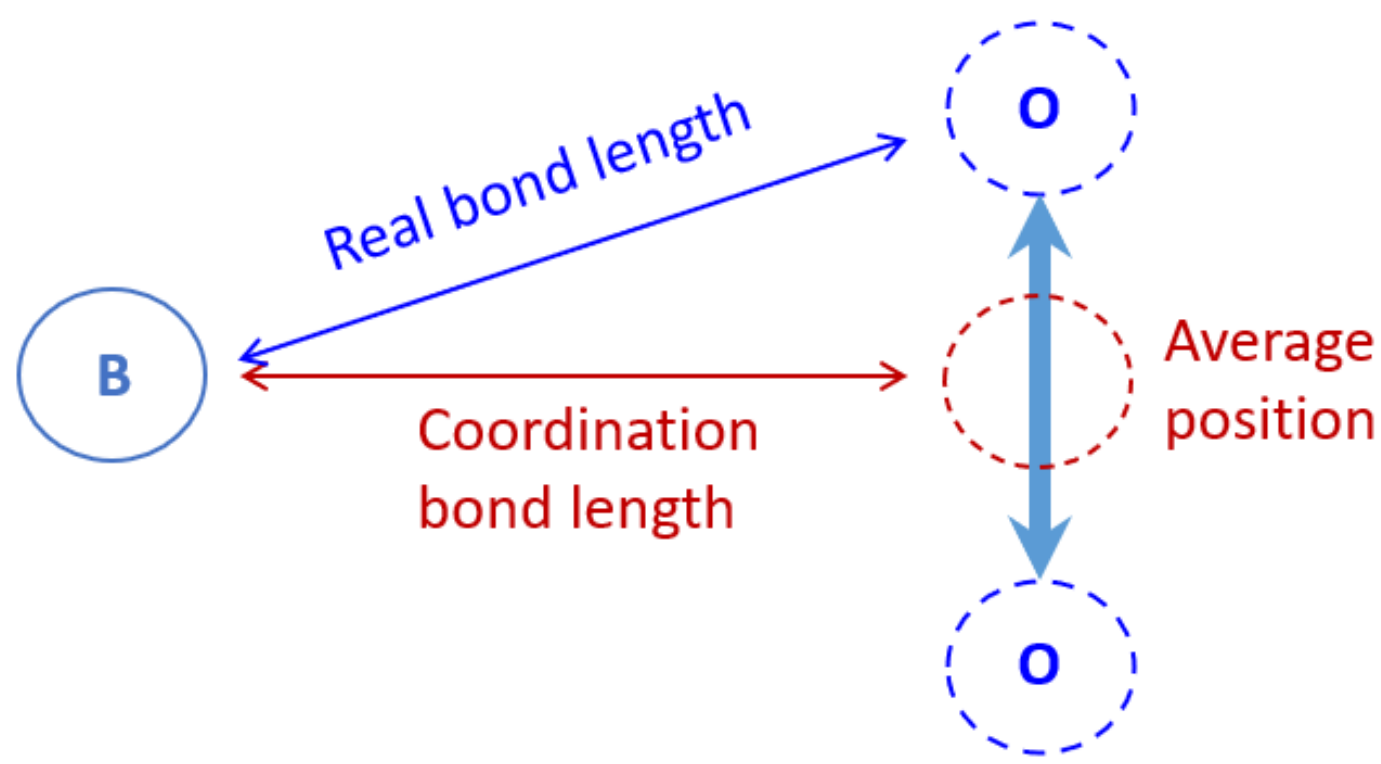

**Figure S4**. Illustration of the distinction between the real bond length and the coordination bond length under conditions of significant transverse vibrations. The coordination bond length refers to the distance between the average atomic positions, whereas the real bond length represents the actual instantaneous distance between atoms.

**Results of Bond valence sum (BVS) calculation**

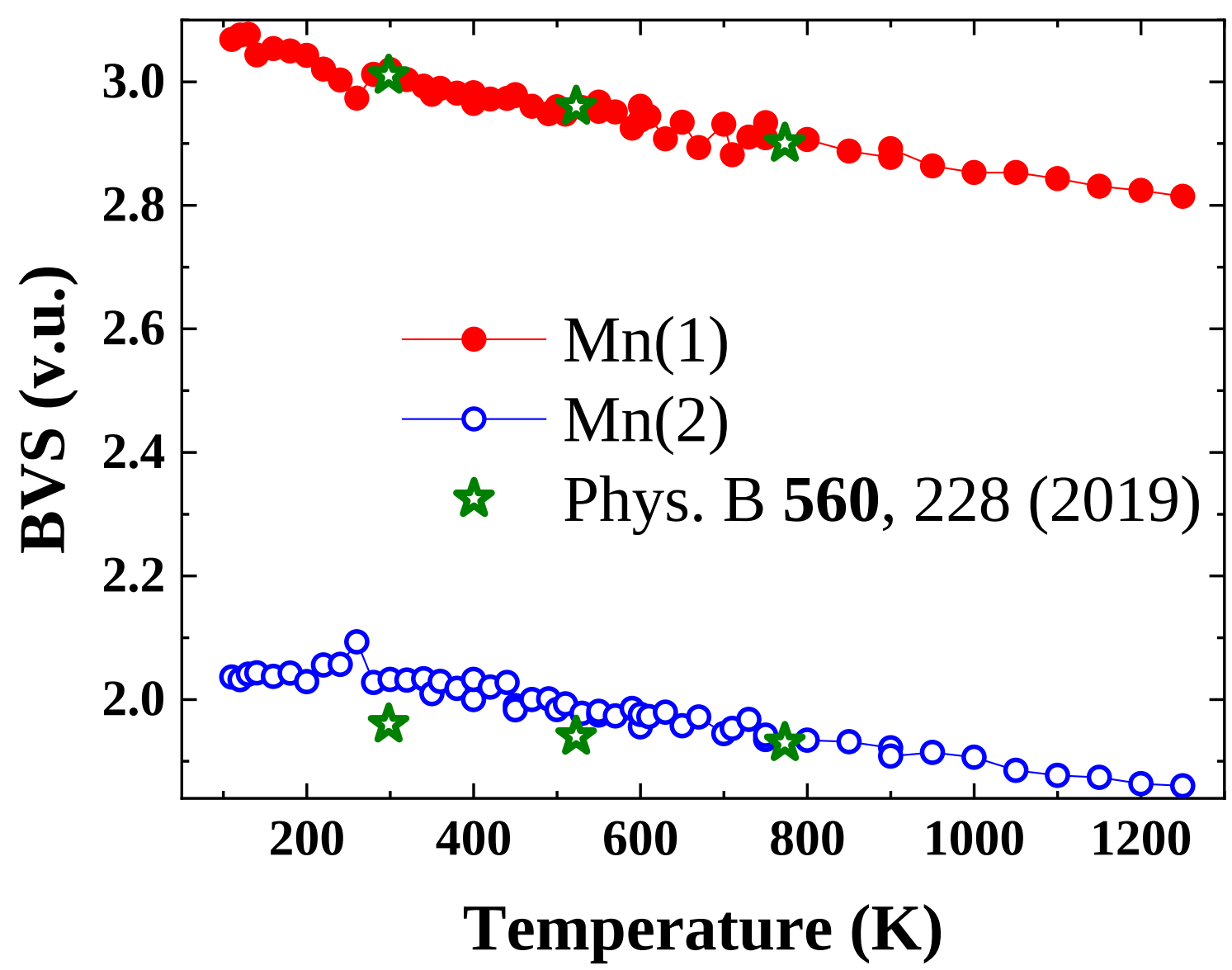

**Figure S5**. Temperature dependence of the BVS. No mix-valence, such as $Mn^{2.5+}$, was observed. The open stars represent the BVS reported by Kazak *et al*. (Phys. B **560**, 228 (2019).) Here, the bond-valence parameters used in the calculation are 1.79 and 1.76 Å for $Mn^{2+}$ and $Mn^{3+}$, respectively. The decrease of BVS while warming is due to the un-modified of the bond-valence parameters with temperatures.